\documentclass[11pt]{article}
\usepackage{moriond,epsfig}
\usepackage{psfrag}
\def\MyPhoto{\psfig{figure=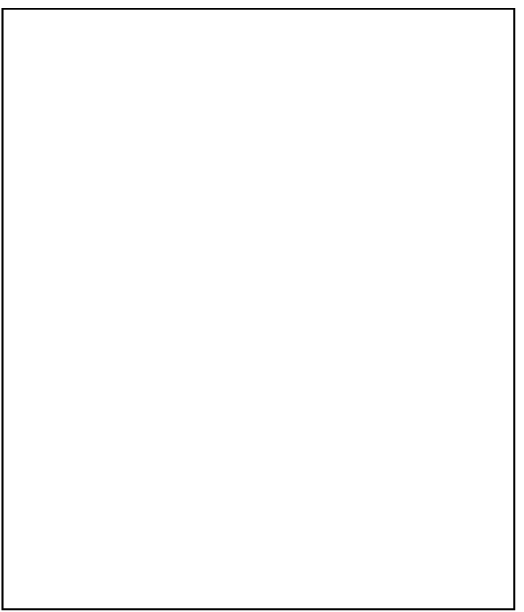,width=35mm}}

\def\be{\begin{equation}}
\def\ee{\end{equation}}
\def\bea{\begin{eqnarray}}
\def\eea{\end{eqnarray}}

\newcommand{\gsim}{\lower .75ex \hbox{$\sim$} \llap{\raise .27ex \hbox{$>$}} }
\newcommand{\lsim}{\lower .75ex \hbox{$\sim$} \llap{\raise .27ex \hbox{$<$}} }

\begin{document}
\vspace*{4cm}
\title{Non-adiabatic instability in coupled dark sectors}

\author{ Laura Lopez Honorez}

\address{Departamento de F\'\i sica Te\'orica \& Instituto de F\'\i sica Te\'orica,\\
Universidad Aut\'onoma de Madrid, 28049 Cantoblanco, Madrid, Spain and\\
 Service de Physique Th\'eorique,
Universit\'e Libre de Bruxelles, 1050 Brussels, Belgium}

\maketitle

\abstracts{ It has been recently pointed out that coupled dark
  matter-dark energy systems suffer from non-adiabatic instabilities at early
  times and  large scales. We show how coupled models free from
  non-adiabatic instabilities can be identified as a function of a
  generic coupling $Q$ and of the dark energy equation of state
  $w$. In our analysis,  we do not refer to any particular cosmic
  field.  We also confront a viable class of model in which the
  interaction  is directly proportional to the dark energy density to
  recent cosmological data. In that framework, we show the
  correlations between  the dark coupling and several cosmological
  parameters allowing to {\it e.g.} larger neutrino mass than
  in uncoupled models.}

\section{Introduction}
\label{subsec:intro}

Interactions between dark matter and dark energy  are still allowed by
observational data today. At the level of the background evolution
equations, one can introduce a coupling
between these two sectors  in the following way:
\begin{eqnarray}
  \label{eq:EOMm}
  \dot\rho_{dm}+ 3\mathcal{H}\rho_{dm} &=& Q\,,\\
\label{eq:EOMe}
 \dot\rho_{de}+ 3 \mathcal{H}\rho_{de}(1+ w)&=&- Q\,,
\end{eqnarray}
where  $\rho_{dm} (\rho_{de})$ denotes the dark matter (dark energy) energy density,  the dot indicates derivative with
respect to conformal time $d\tau = dt/a$, $\mathcal{H}= {\dot a}/a$
and  $w=P_{de}/\rho_{de}$ is the dark-energy equation of state ($P$
denotes the pressure).
We work with the Friedman-Robertson-Walker (FRW) metric,
assuming a flat universe and  pressureless dark matter $w_{dm} =
P_{dm}/\rho_{dm}=0$. 

Q encodes the dark coupling and drives the energy exchange between dark
matter and dark energy. For {\it e.g.} $Q<0$ the energy flows from
dark matter to dark energy. It also changes the way that dark
matter and dark energy redshift acting as an extra
contribution to their effective equation of state. In particular, for
{\it e.g.} $Q<0$, dark matter redshifts faster so that there is  more
dark matter in the past compared to uncoupled
scenarios assuming that the dark matter
density today is the same in the two models. Matter-radiation equality happens earlier 
and the growth of dark matter clustering is enhanced.
This is one of the features which enable us to constraint the model
with available cosmological data (see section~\ref{sec:rhoe}).

In order to deduce the evolution of  density and velocity perturbations in  coupled models, we need an
expression of the energy transfer in terms of the stress-energy tensor.
We follow Ref.\cite{Valiviita:2008iv} in parameterizing the interaction as:
 \begin{eqnarray}
  \label{eq:conservDM}
\nabla_\mu T^\mu_{(dm)\nu} &=&Q \,u_{\nu}^{(dm)}/a~, \\
  \label{eq:conservDE}
\nabla_\mu T^\mu_{(de)\nu} &=&-Q \,u_{\nu}^{(dm)}/a~, 
\end{eqnarray}
with $T^\mu_{(dm)\nu}$ and $T^\mu_{(de)\nu}$ the energy momentum tensors for the dark matter 
and dark energy components, respectively. The dark matter four velocity $u_{\nu}^{(dm)}$ is defined 
in the synchronous gauge, in terms of the fluid proper velocity $v^i_{(dm)}$, as 
$u^{(dm)}_\nu = a (-1, v^i_{(dm)})$, where $\mu=0..3$ and $i=1..3$.
This choice of parameterization guaranties the conservation of the
total stress energy tensor of the system while it
 avoids momentum transfer in the rest frame of dark 
matter in which case one can work in the synchronous gauge comoving with dark matter 
({\it i.e} $v^i_{(dm)}=0$).

We provide~\cite{Gavela:2009cy} a criteria associated to the dubbed {\bf \it the doom factor}
(see section \ref{sec:doom})  to identify the stability region
of coupled models satisfying to Eq.~(\ref{eq:conservDM})
and~(\ref{eq:conservDE}).
 The doom factor is a function of the model
parameters such as $Q$ and $w$ but it is
defined independently of the explicit form of the coupling $Q$. Notice
that the dark coupling terms which appears in the non-adiabatic dark
 energy pressure perturbations were first  pointed
 out as a source for  early time
 instabilities  at large scales in Ref.~\cite{Valiviita:2008iv}.

Our results will then be illustrated within a successful class of models, in which $Q$ is proportional 
to the dark energy density. Present data will be shown to allow for a sizeable interaction strength and to imply weaker cosmological limits  on 
neutrino masses with respect to non-interacting
scenarios. 
\section{Origin of non-adiabatic instabilities}
\label{sec:instab}

Non-adiabatic instabilities arises at linear order in perturbations.
Using the publicly available CAMB code~\cite{Lewis:1999bs}, we
preferred to work in the synchronous gauge comoving with dark matter.

\subsection{Doom factor}
\label{sec:doom}
 Given our gauge choice, it is
necessary to work out the expression for dark energy pressure perturbations
$\delta P_{de}$ in the rest frame for dark matter. It can be
shown that in the presence of a dark coupling (see {\it e.g.} Ref.~\cite{Kodama:1985bj} and  also
Ref.\cite{Valiviita:2008iv}) it is given by:
\begin{eqnarray}
\frac{\delta P_{de}}{\delta \rho_{de}}
 &=&\hat c_{s\,de}^2 +3(\hat c_{s\,de}^2- c_{a\,de}^2) 
 (1+w)\left(1 + {\bf d}  \right) \frac{{\mathcal H}\theta_{de}}{k^2\delta_{de}}\,,
 \label{eq:dpcs}
\end{eqnarray}
where $\delta \rho_{de}$ denotes the dark energy energy density
perturbation and $\delta_{de}=\delta \rho_{de}/\rho_{de}$,  $\theta_{de} \equiv
\partial_i v^i_{(de)}$ is the divergence of the fluid proper velocity $v^i_{(de)}$,
$\hat c_{s\,de}^2$  is the propagation speed of pressure fluctuations in the
rest frame of dark energy and $c_{a\,de}^2= \dot P_{de}/\dot \rho_{de}$ is the so called
``adiabatic sound speed''. 
In the following, we work with constant equation of state $w$ in which case
$c_{a\,de}^2=w$ and we assume that our universe is  in accelerating expansion today which
implies that $w<-1/3$. Moreover we restrict our analysis to the case  $\hat
c_{s\,de}^2>0$ and $\hat
c_{s\,de}^2=1$ will be assumed for numerical computation.

In Eq.~(\ref{eq:dpcs})  ${\bf d}$ refers to the doom factor which we
have defined as:
\begin{equation}
\label{eq:maldito}
 {\bf d} \equiv \frac{Q}{3\mathcal H\rho_{de}(1+w)}\,.
\end{equation}
We dub it so as it is precisely this extra factor, proportional to the dark coupling $Q$, which may induce non-adiabatic instabilities in the 
evolution of dark energy perturbations. 
Its sign will be determinant, as we show in the next section.
\subsection{Growth equation}
\label{sec:growth}
A cartoon equation of the growth equation governing the evolution of  energy density linear
perturbation for any species $i, j$ is given by:  
\begin{center}
\vspace*{-2cm}
\rotatebox{270}{\includegraphics[width=0.40\textwidth]{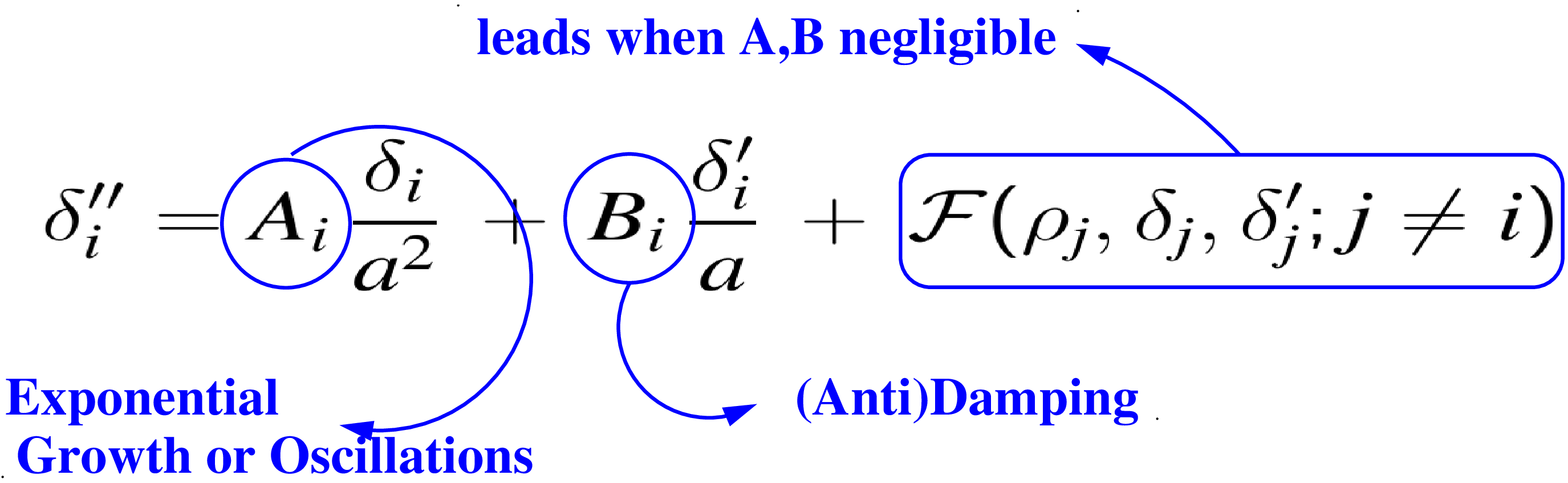}}
\end{center}

\vspace*{-1.5cm}
where $\delta_i= \delta \rho_i/\rho_i$ and $'\,= \partial/\partial \,a$.
The evolution of a perturbation depends on the relative weight of the
three terms present in the equation {\it and} on their signs: 
\begin{enumerate}
\item For positive $A$, the $A$ and $B$ terms taken by themselves would induce a rapid growth of 
the perturbation, which may be damped or antidamped (reinforced) depending on whether $B$ is 
negative or positive, respectively~\footnote{Obviously, for $|B|>>|A|$ a negative $B$ would prevent 
the onset of growth for any sign of $A$.}. In particular, for $A$ and $B$ both positive, the solution 
may enter in an exponentially growing, unstable, regime. \label{un}
\item For negative $A$, in contrast, the $A$ and $B$ terms taken alone describe a harmonic oscillator, 
with oscillations damped (antidamped) if $B$ is negative (positive). In the $A,B<0$ regime, the third 
term may plays in fact the leading role. \label{dos}
  \end{enumerate}

In the standard  uncoupled scenario  dark matter perturbations behave as in case \ref{un} above (with $A>0$ and $B<0$), 
while dark energy ones provide an example of behavior as in case \ref{dos}.
For coupled models,  we concentrate on the case in which the dark-coupling terms
dominate over the usual one in order to put forward the presence of non
adiabatic instabilities (see Ref.\cite{Gavela:2009cy} for more details). This
{\it strong coupling regime} can be characterized by 
 \begin{eqnarray}
 |{\bf d}|&=&
\left|\frac{Q}{3\mathcal H\rho_{de}(1+w)}\right|\,>\,1 \,,
  \label{eq:condstr}
\end{eqnarray}
which also guarantees that the interaction among the two dark sectors drives the non-adiabatic 
contribution to the dark energy pressure wave, see Eq.~(\ref{eq:dpcs}).  At
large scales and early times, it can be worked out that the main
contributions of $\delta_{de}$
and $\delta_{de}'$ coefficients  to the second order differential equation reduce to:
  \begin{eqnarray}
  \label{eq:grstrongfullours}
\delta_{de}''&\simeq&  \,3\,{\bf d}\,(\hat c_{s\,de}^2+1)\left(\,\frac{\delta_{de}'}{a} 
    \,+\,3\frac{\delta_{de}}{a^2}\frac{(\hat c_{s\,de}^2-w)}{\hat c_{s\,de}^2+1}
     \,+\,\frac {3(1+w)}{a^2}\delta[\,{\bf d}\,]\,\right)+...
  \end{eqnarray}
The sign of the  coefficient $B_e$ of $\delta_{de}^\prime$  in this expression is crucial for 
the analysis of instabilities.  Assuming $\hat c_{s de}^2>0$, it reduces to the sign of the doom 
factor ${\bf d}$ defined in Eq.~(\ref{eq:maldito}).

Similar second order differential equations for $\delta_{de}$ were obtained in
Ref.~\cite{He:2008si} for particular expressions of the dark coupling $Q$ and an analytical form of their solutions where
derived in order to determine when $\delta_{de}$ blows up. In particular, the results of Ref.~\cite{He:2008si}
confirm those of Ref.~\cite{Valiviita:2008iv}  for  positive $Q\propto
\rho_{dm}$ and $1+w>0$.  In comparison, our approach gives  
rise to general conditions to avoid instabilities as a function of the sign of the doom factor
independently of the exact form of the dark coupling $Q$. Indeed, as
previously argued, a positive  ${\bf d}$  acts as an antidamping source in the growth 
Eq.~(\ref{eq:grstrongfullours}). Whenever ${\bf d}>1$,  the overall sign of the $A_e$ coefficient of $\delta_{de}$, resulting from the 
last two terms in Eq.~(\ref{eq:grstrongfullours}), is  also positive
and it triggers an exponential runaway 
growth of the dark energy perturbations.
 Large scale instabilities arise then and the universe appears to be non viable.

\section{A viable model: $Q\propto\rho_{de}$ }
\label{sec:rhoe}

We have developed a method to determine if a coupled model
satisfying Eq.~(\ref{eq:conservDM}) and~(\ref{eq:conservDE}) suffer or
not from non-adiabatic instabilities. We can now easily verify that for
\begin{equation}
  Q=\xi {\mathcal H} \rho_{de},
\label{eq:Qrhoe}
\end{equation}
we have a rather  simple and  viable model for specific
combination of $1+w$ and of the dimensionless constant
coupling $\xi$.
In this model, the doom factor of Eq.~(\ref{eq:maldito}) is  given  by:
 \begin{equation}
\label{eq:maldito_us}
{\bf d }= \frac{\xi}{3(1+w)}\,.
\end{equation}
Its sign defines the (un)stable regimes.   When  ${\bf d } <0$, that
is, for $\xi<0$ and $1+w>0$ (or $\xi>0$ and $1+w<0$), no instabilities
are expected. On the contrary, when $\xi$ and $1+w$ have the same sign, instabilities will
  develop at early times whenever ${\bf d } >1$ .

In Ref.~\cite{Gavela:2009cy}  we confirmed these results
numerically. Also notice that they are in agreement with those of
Ref.~\cite{He:2008si}, which first pointed out that  coupled models
with $Q=\xi {\mathcal H} \rho_{de}$ can be stable for $1+w <0$.
They restricted though  their  stability analysis to the $\xi>0$
case.

In the following, we confront the model satisfying to Eq.~(\ref{eq:Qrhoe}) to cosmological data restricting
ourselves  to negative couplings and $w>-1$. This guarantees that instability problems in the dark energy perturbation equations are
  avoided for all values\footnote{For $Q=\xi {\mathcal H}
    \rho_{de}$, the dark energy density is always positive, all along
    the cosmic evolution and since its  initial moment. To ensure that
    the same happens with the dark matter density, all values of 
 $w<0$ are acceptable for $\xi<0$, while for positive $\xi$ it is 
required that $\xi\lsim -w$.}  of $\xi$.

\section{Cosmological Constraints from data for $Q=\xi {\mathcal H} \rho_{de}$}
\label{sec:cosmo}
%

%
%%%%%%%%%%%%%%%%%%%%%%%%%%%%%%%%%%%%%%%%%%%%%%%%%%%%%%%%%%%%%%%%%%%%%
\begin{figure}[t]
\vspace{-0.1cm}
\begin{center}
\begin{tabular}{cc}
\hspace*{-0.75cm} 
%\psfrag{w}[c][c]{{$w$}}
\includegraphics[width=8cm]{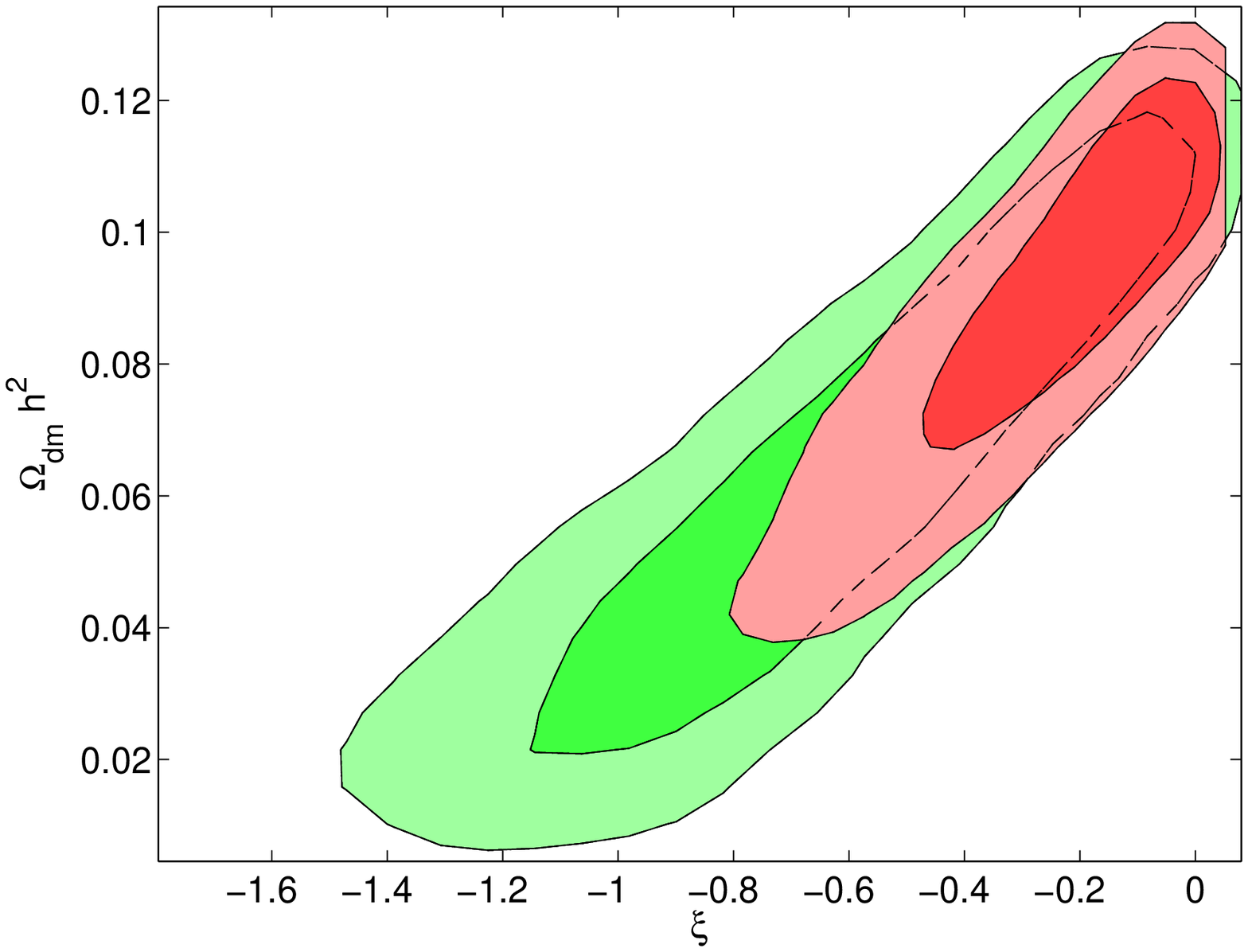} &
\includegraphics[width=8cm]{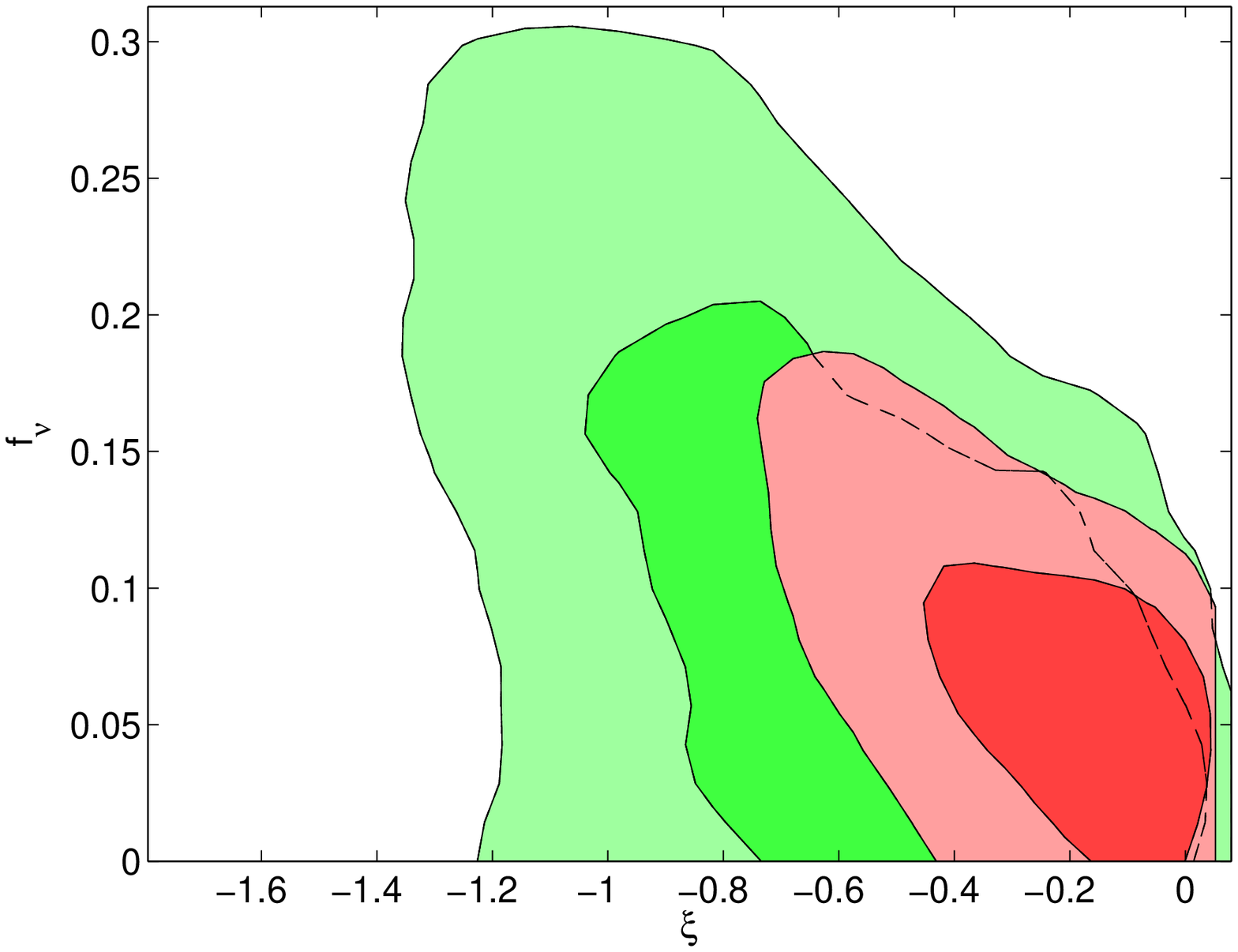} \\
\end{tabular}
\caption{\it Scenario with $Q\propto \rho_{de}$. Left (right) panel: 1$\sigma$ and 2$\sigma$ 
marginalized contours in the $\xi$--$\Omega_{dm} h^2$  ($\xi$--$f_\nu$) plane. The largest, green contours show 
the current constraints from WMAP (5 year data), HST, SN and $H(z)$ data. The smallest, red contours 
show the current constraints from WMAP (5 year data), HST, SN, $H(z)$ and LSS data.}
\label{fig:fig1o}
\end{center}
\end{figure}

%%%%%%%%%%%%%%%%%%%%%%%%%%%%%%%%%%%%%%%%%%%%%%%%%%%%%%%%%%%%%%%%%%%%%%

We explored  the current constraints on the dark
energy-dark matter coupling $\xi$ using the publicly available package
\texttt{cosmomc}~\cite{Lewis:2002ah}. The latter was  modified in order to
include the coupling among the dark matter and dark energy
components. More details on the cosmological model and
on the priors adopted can be found in Ref.~\cite{Gavela:2009cy}.
The datasets which were taken into account in the analysis are:
\begin{enumerate}
\item WMAP 5-year data~\cite{Dunkley:2008ie,Komatsu:2008hk}
\item  prior on
the Hubble parameter of $72\pm8$~km/s/Mpc from the Hubble key
project (HST)~\cite{Freedman:2000cf}
\item  Super Novae (SN) data~\cite{Kowalski:2008ez}
\item $H(z)$ data at $0<z<1.8$ from galaxy ages~\cite{Simon:2004tf}
\item large scale structure data (LSS data)  from the Sloan Digital Sky
Survey~\cite{Tegmark:2006az}
\end{enumerate}
The data analysis was carried out into two runs,  the first run
includes the datasets from 1 to 4 while in the second run the fifth
dataset  is added.

Figure~\ref{fig:fig1o} (left panel) illustrates the $1$ and $2\sigma$
marginalized contours in the $\xi$--$\Omega_{dm} h^2$ plane where
$\Omega_{dm}$ is today's ratio between dark matter energy density and
critical energy density. 
The results from the two runs described above are shown.
Notice that a huge degeneracy is present, being $\xi$ and $\Omega_{dm}
h^2$ positively correlated.
 The shape of the contours can be easily
understood following our discussion in section~\ref{subsec:intro}. In
a universe with a negative dark coupling $\xi$, dark matter redshift
faster. As a consequence, the matter content in the past  is higher
than in the standard uncoupled scenario for a fixed dark matter
density today. The amount of \emph{intrinsic} dark
matter (which is directly proportional to $ \Omega_{dm} h^2$) needed
to reproduce  the LSS data should decrease as the dark coupling becomes more and
more negative. We also see that LSS data in the second run (red
contours in Fig.~\ref{fig:fig1o})  give rise to
the most stringent constraint on the coupling $\xi$.

The right panel of Fig.~\ref{fig:fig1o} shows  the correlation
among the fraction of matter energy-density in the form of massive
neutrinos $f_\nu$ and the dark coupling $\xi$. The relation between
the neutrino fraction used here $f_\nu$ and the neutrino mass for $N_\nu$ degenerate neutrinos reads
\begin{equation}
f_\nu=\frac{\Omega_\nu h^2}{\Omega_{dm} h^2}=\frac{\sum m_\nu}{93.2 \textrm{eV}} \cdot \frac{1}{\Omega_{dm} h^2}=\frac{N_\nu m_\nu}{93.2 \textrm{eV}}\cdot \frac{1}{\Omega_{dm} h^2} ~.
\end{equation}
Neutrinos can indeed play a relevant role in large scale structure
formation and leave key signatures in several cosmological data sets,
Specially, non-relativistic neutrinos in the
recent Universe suppress the growth of matter density fluctuations and
galaxy clustering. This effect can be compensated  by the existence of
a coupling between the coupled sectors, given that in this model
negative couplings enhance the growth of matter density perturbations.

\section{Conclusion}
\label{sec:concl}
In this talk, we discuss the origin of non-adiabatic instabilities in
coupled models satisfying to
Eqs.~(\ref{eq:conservDM}) and~(\ref{eq:conservDE}). We show that the
sign of the doom factor ${\bf d}$
\begin{equation}
{\bf d} \equiv \frac{Q}{3\mathcal H\rho_{de}(1+w)}
\end{equation}
 which is a function of the dark coupling $Q$ characterizes the (un)stable regime. 
In particular, when ${\bf d }$ is positive and sizeable, ${\bf d }>1$, the dark-coupling 
dependent terms may dominate the evolution of dark energy perturbations, which will then enter 
a runaway, unstable, exponential growth regime.

 In a class of viable model in which $Q=\xi {\mathcal H} \rho_{de}$ we have then
 studied the constraints from cosmological data on the dimensionless
 coupling $\xi$. This analysis was carried out in the  $\xi<0$ and positive 
$(1+w)$ region of the parameter space which offers the best agreement with data on large scale 
structure formation. Both $w$ and $\xi$ are not very
constrained from data, and it can be shown~\cite{Gavela:2009cy} that substantial values 
for both parameters, near -0.5, are easily allowed. Furthermore, 
$\xi$ turns out to be positively correlated with  $\Omega_{dm}
h^2$ and larger neutrino fraction $f_{\nu}$ is allowed for negative $\xi$.

\section*{Acknowledgments}
The work reported has been done in collaboration with B.~Gavela,
D.~Hernandez, O.~Mena and S.~Rigolin and  the author was partially supported by CICYT through
the project FPA2006-05423, by CAM through the project HEPHACOS, P-ESP-00346,  by the PAU (Physics
of the accelerating universe) Consolider Ingenio 2010, by the
F.N.R.S. and  the I.I.S.N..
\section*{References}
%% \begin{thebibliography}{99}
%% \bibitem{ja}C Jarlskog in {\em CP Violation}, ed. C Jarlskog
%% (World Scientific, Singapore, 1988).

%% \bibitem{ma}L. Maiani, \Journal{\PLB}{62}{183}{1976}.

%% \bibitem{bu}J.D. Bjorken and I. Dunietz, \Journal{\PRD}{36}{2109}{1987}.

%% \bibitem{bd}C.D. Buchanan {\it et al}, \Journal{\PRD}{45}{4088}{1992}.

%% \end{thebibliography}
\bibliographystyle{unsrt} 
\bibliography{bibdmde-v3.bib}

\end{document}